\documentclass[12pt]{article}
\usepackage{hhline,array,amsmath,graphicx}
\usepackage{pifont,latexsym,ifthen,theorem,rotating,calc}
\usepackage{siunitx}
\usepackage{etoolbox}
\usepackage{mathtools}
\usepackage{amsmath}
\usepackage{bbm}
\usepackage{pdflscape}
\usepackage{rotating}
\usepackage{bm}
\usepackage{enumitem}

\title{\textbf{A Linear Relationship between Correlation and Cohen's Kappa for Binary Data and Simulating Multivariate Nominal and Ordinal Data with Specified Kappa Matrix}}

\author{Soumya Sahu\thanks{Soumya Sahu (e-mail:ssahu6@uic.edu) is PhD student of Biostatistics, and Hakan Demirtas is Associate Professor of Biostatistics, Division of Epidemiology and Biostatistics (MC923), University of Illinois at Chicago, 1603 West Taylor Street, Chicago, IL, 60612.} \hspace{0.1cm} and Hakan Demirtas}

\begin{document}
\date{\today}
\baselineskip=22.5pt
\maketitle

\begin{abstract}

\noindent Cohen's kappa is a useful measure for agreement between the judges, inter-rater reliability, and also goodness of fit in classification problems. For binary nominal and ordinal data, kappa and correlation are equally applicable. We have found a linear relationship between correlation and kappa for binary data. Exact bounds of kappa are much more important as kappa can be only .5 even if there is very strong agreement. The exact upper bound was developed by Cohen (1960) but the exact lower bound is also important if the range of kappa is small for some marginals. We have developed an algorithm to find the exact lower bound given marginal proportions. Our final contribution is a method to generate multivariate nominal and ordinal data with a specified kappa matrix based on the rearrangement of independently generated marginal data to a multidimensional contingency table, where cell counts are found by solving system of linear equations for positive roots.
\\

\noindent
KEY WORDS: Cohen's Kappa; Correlation; Random number generation
\end{abstract}

\section{Introduction}

In clinical studies, it frequently occurs that useful level of measurement is obtainable only in nominal and ordinal scale. The reliability of these measurements is determined by the degree, significance, and sampling stability of the agreement between the raters or judges who independently rate the sample of units into nominal or ordinal categories (Jacob Cohen, 1960). In healthcare industries, study designs involve measuring the extend to which data collectors record the same scores for the same phenomena and study results is partly a function of amount of disagreement among the data collectors (Mary L.McHugh, 2012). So, it is essential to measure the agreement among data collectors, which is called "interrater reliability". In classification problems, goodness of fit of a particular machine-learning method is judged by the measure of agreement between the observed categories and the categories determined by that method. Most importantly, in goodness of fit measurement these categories are treated as nominal even if they are ordered. The above scenarios motivate the need of a good measure of agreement between the raters for nominal and ordinal data. One primitive measure is sum of the proportion of cases in which the raters agreed. Cohen (1960) criticized this approach as it does not account for the amount of agreement is to be expected by chance and suggested a measure which is famously known as Cohen's Kappa. Let us consider $K$ mutually exclusive categories of a nominal or ordinal variable. Two raters A and B classify $N$ objects to these categories with marginal proportions $p_1^A, p_2^A,...,p_K^A$ and $p_1^B, p_2^B,...,p_K^B$ respectively. Let, $p_0$ be the proportion of the cases where both the raters agreed and $p_c$ be the proportion of the cases where agreement is expected by chance; $p_c$ can be written as $\sum \limits_{i = 1}^K p_i^A p_i^B$. Cohen's kappa is defined as, $$\kappa = \frac{p_0 - p_c}{1 - p_c}.$$
William A. Scott (1955) proposed a similar measure, Pi, where he defined $p_c$ as squared arithmetic mean of marginal proportions ($p_c^{Scott} = \sum \limits_{i = 1}^K {(\frac{p_i^A + p_i^B}{2})}^2$), where $\kappa$ uses squared geometric mean. Scott's Pi assumes same distribution of responses for each raters, which makes $\kappa$ more general and informative. Joseph L. Fleiss (1971) extended the Pi coefficient to accommodate multiple raters. In this paper we haven't consider any single measure of agreement for multiple raters (i.e. the agreement implies same response from all the raters), whereas, we have discussed pairwise $\kappa$ measure for multiple raters and we have considered a kappa matrix which is structurally similar as correlation matrix.\\

Pearson product moment correlation coefficient is not an appropriate measure of association for nominal data. For example, consider two raters and they have rated in nominal categories 1 to 6, if there are all agreements in categories 1, 6 ((1, 1), (6, 6) pairs occurred with probability .4 each) than all disagreement categories 2, 3, 4, 5 ((2, 5), (3, 4), (4, 3), (5, 2) pairs occurred with probability .05 each), resulting correlation (.905) will be positive and close to 1. In the opposite situation i.e. if there are all agreements in categories 3, 4 ((3, 3), (4, 4) pairs occurred with probability .4 each) than all disagreement categories 1, 2, 5, 6 ((1, 6), (2, 5), (5, 2), (6, 1) pairs occurred with probability .05 each), resulting correlation (-.62) will be negative and close to -1. This should not be a desired property of an appropriate association measure in nominal scale because one can not claim that 1 $<$ 3 or 6 $>$ 4 for nominal variables. $\kappa$ measure is appropriate in this scenario as it does not account for any ordering in the measurement scale. In both of the above situations $\kappa$ is same (.7). In this point of view, it may not be possible to find any relationship between correlation and kappa when there are more than two categories in measurement scale. When there are only two categories, the above situation can not occur. So, both of correlation and $\kappa$ are appropriate association measure for binary nominal and ordinal data.  We have found a linear relationship between $\kappa$ and correlation for binary data. (see section 2)\\

Cohen (1960) showed that $\kappa$ can vary between -1 and +1 depending on the marginal probability distribution of raters.
A disadvantage of using $\kappa$ measure is that it may not take value close to 1 when there is proportion of agreement close to 1.
McHugh (2012) mentioned an example where the problem of interest was to judge the agreement between human and automated rater and in spite of having 94.2 \% agreement, $\kappa$ value was only .555. So, observing only $\kappa$ measure it very difficult to come to any conclusion that automated raters are reliable or not. A useful thing in this context may be to find the exact lower and upper bounds of $\kappa$ for obtained marginal probabilities of raters and compare the observed $\kappa$ with those bounds. Cohen (1960) mentioned the exact upper bound can be found by replacing $p_0$ by $p_0^M$, where $p_0^M$ is found by pairing the $p_i^A$ and $p_i^B$ values, selecting the smaller of each pair, and summing the $K$ values. The lower bound is also important to understand the strength of observed $\kappa$ by finding how closer it is to the upper bound than the lower bound. In section 4, we shall discuss our proposed algorithm to find the exact lower bound of $\kappa$ given marginal probabilities.\\

Although, there are several tools to generate multivariate ordinal data with specified correlation matrix and marginal probabilities (Demirtas 2006, Amatya \& Demirtas 2015) but there is absence of appropriate tool to generate bivariate or multivariate nominal data. Demirtas (2019) has developed a sorting based approach to generate bivariate data with any marginals for specified feasible correlation. As we have developed a linear relationship between $\kappa$ and correlation for binary data, Demirtas 2019 can be applicable to generate bivariate binary nominal or ordinal data with specified marginals and $\kappa$. Unfortunately, it fails to do the same when there are more than two categories per raters (see section 3). We have developed a method to generate bivariate nominal or ordinal data with specified feasible $\kappa$ with any number of categories for marginal distributions. The method involves filling up a $K \times K$ table (2 dimensional array) after generating two marginals independently with specified marginal probabilities. Once marginals are generated $p_c$ is fixed, so we can adjust $p_0$ by playing with the diagonal entries to achieve desired $\kappa$. This approach has been generalized to multivariate case by using multidimensional array. We have discussed these approaches elaborately in section 3 and 5. 

\section{A linear relationship between correlation ($\rho$) and Cohen's kappa ($\kappa$) for binary data}
\subsection{The linear relationship}
Consider two binary variables $X$ and $Y$ with marginal probabilities $(1-p_1, p_1)$ and $(1-p_2, p_2)$
respectively. For the given value of correlation, $\rho$, the joint distribution of $X$ and $Y$ are fully specified, i.e. we shall be able to calculate cell probabilities as a function of $p_1$, $p_2$ and $\rho$. Cohen's kappa, $\kappa$, is a function of cell probabilities only, so, it must a function of $p_1$, $p_2$ and $\rho$. Consider the following table as the joint distribution of $X$ and $Y$,
\begin{center}
    \begin{tabular}{|c|c|c|c|}
\hline 
 & $Y = 0$ & $Y = 1$ & Marginal of $X$ \\ 
\hline 
$X = 0$ & $d$ & $c$ & $c + d$ $=$ $1 - p_1$ \\ 
\hline 
$X = 1$ & $b$ & $a$ & $a + b$ $=$ $p_1$\\ 
\hline 
Marginal of $Y$ & $b + d$ $=$ $1 - p_2$ & $a + c$ $=$ $p_2$ & 1 \\ 
\hline 
\end{tabular} 
\end{center}

By definition, $\kappa = \frac{p_0 - p_c}{1 - p_c},$ where, $p_c = p_1p_2 + (1-p_1)(1-p_2).$ We need to show $p_0$ as a function of $p_1$, $p_2$ and $\rho$. \\
$p_1 = a+b$, $p_2 = a+c \implies p_1 - (1-p_2) = a-d \implies d = a + (1-p_1-p_2).$\\
$\rho = \frac{a - p_1p_2}{\sqrt{(1-p_1)(1-p_2)p_1p_2}} \implies a = \rho\sqrt{(1-p_1)(1-p_2)p_1p_2} + p_1p_2.$\\
So, $d = \rho\sqrt{(1-p_1)(1-p_2)p_1p_2} + (1-p_1)(1-p_2)$ and $p_0 = a+d = 2\rho\sqrt{(1-p_1)(1-p_2)p_1p_2} + p_1p_2 + (1-p_1)(1-p_2).$\\
Finally, $\kappa = \frac{2\rho\sqrt{(1-p_1)(1-p_2)p_1p_2}}{1-p_1p_2 - (1-p_1)(1-p_2)} = \rho C,$ where, $C = \frac{\sqrt{(1-p_1)(1-p_2)p_1p_2}}{1-p_1p_2 - (1-p_1)(1-p_2)}.$\\

\textbf{Claim:} $0 < C \leq 1$, equality holds iff $p_1 = p_2,$ given $0 < p_1 < 1, 0 < p_2 < 1.$\\
\textbf{Proof:} Assume, $C > 1$. This implies, $2\sqrt{p_1p_2q_1q_2} > 1 - p_1p_2 - q_1q_2 \implies (\sqrt{p_1p_2} + \sqrt{q_1q_2})^2 > 1 \implies \sqrt{p_1p_2} + \sqrt{q_1q_2} > 1 \implies p_1p_2 > (1-\sqrt(q_1q_2))^2 \implies (1-q_1)(1-q_2) > 1 + q_1q_2 - 2\sqrt{q_1q_2} \implies (\sqrt{q_1} - \sqrt{q_2})^2 < 0,$ which is false.\\
Here, $q_1 = 1-p_1, q_2 = 1-p_2.$ Equality holds iff $(\sqrt{q_1} - \sqrt{q_2})^2 = 0 \implies p_1 = p_2.$\\
Using Arithmetic Mean $\geq$ Geometric Mean for positive quantities, $\frac{p_1 + p_2}{2} \geq \sqrt{p_1p_2} > p_1p_2,$ as $0 < p_1 < 1, 0 < p_2 < 1.$
So, $p_1 + p_2 - 2p_1p_2 > 0 \implies 1-p_1p_2 - (1-p_1)(1-p_2) > 0 \implies C > 0.$\\

The above claim proves that correlation is greater than or equals to kappa in absolute value i.e. $|\rho| \geq |\kappa|,$ for binary data with positive marginal probabilities.
So, the relationship we found between $\rho$ and $\kappa$ is not only linear but also passing through the origin and having a positive slope which is less than equals to 1 in absolute value.

\subsection{Bounds of $\kappa$ for binary case}
For bivariate binary data with probabilities $(p_1, q_1)$, $(p_2, q_2)$, the correlation is bounded below by $max\{-\sqrt{\frac{p_1p_2}{q_1q_2}}, -\sqrt{\frac{q_1q_2}{p_1p_2}}\}$ and bounded above by $min\{\sqrt{\frac{p_1q_2}{q_1p_2}}$, $\sqrt{\frac{p_2q_1}{p_1q_2}}\}$ (Emrich and Piedmonte 1991). Using the linear relationship we have found in section 2.1, we can easily claim that, under the same set up, $\kappa$ is bounded below by $max\{-\frac{2p_1p_2}{1-p_1p_2 - q_1q_2}, -\frac{2q_1q_2}{1-p_1p_2 - q_1q_2}\}$ and bounded above by $min\{\frac{2p_1q_2}{1-p_1p_2 - q_1q_2}$, $\frac{2p_2q_1}{1-p_1p_2 - q_1q_2}\}$.
 (obtained by multiplying correlation bounds by $C$, as $C > 0.$)\\
 
As we have discussed in section 1, Cohen (1960) found the upper bound of $\kappa$ by replacing $p_0$ by $\sum \limits_{i = 1}^K min\{p_i^A, p_i^B\}$ in the definition of $\kappa$. So, using this, upper bound of $\kappa$ in binary case is,\\
if $p_1 > p_2$, $\frac{p_2 + q_1 - p_1p_2 - q_1q_2}{1-p_1p_2 - q_1q_2} = \frac{p_2(1-p_1) + q_1(1 - q_2)}{1-p_1p_2 - q_1q_2} = \frac{2p_2q_1}{1-p_1p_2 - q_1q_2},$\\
if $p_1 < p_2$, $\frac{p_1 + q_2 - p_1p_2 - q_1q_2}{1-p_1p_2 - q_1q_2} = \frac{p_1(1-p_2) + q_2(1 - q_1)}{1-p_1p_2 - q_1q_2} = \frac{2p_1q_2}{1-p_1p_2 - q_1q_2},$\\
if $p_1 = p_2$, $\frac{p_1 + q_1 - p_1^2 - q_1^2}{1-p_1p_2 - q_1q_2} = \frac{p_1(1-p_1) + q_1(1 - q_1)}{1-p_1p_2 - q_1q_2} = \frac{2p_1q_1}{1-p_1p_2 - q_1q_2}.$\\
Notice that these upper bounds are exactly same as the upper bounds we found in previous paragraph, because,\\
when $p_1 > p_2$, $min\{p_1q_2, p_2q_1\} = p_2q_1$, when $p_1 < p_2$, $min\{p_1q_2, p_2q_1\} = p_1q_2$ and when $p_1 = p_2$, $p_1q_2 = p_2q_1 = p_1q_1.$\\
By putting $p_0 = 0$ in definition of $\kappa$ we can obtain the lower bound but $p_0 = 0$ is not feasible for some choices of marginal probabilities, for example, $p_1 = 0.8$ and $p_2 = 0.7$. $p_0 = 0$ is feasible iff $p_1 + p_2 = 1$, so, under this condition, lower bound of $\kappa$ is $\frac{0 - p_1p_2 - q_1q_2}{1 - p_1p_2 - q_1q_2} = \frac{-2p_1p_2}{1 - p_1p_2 - q_1q_2}$ as $p_1 = q_2$ and $p_2 = q_1.$ Notice that this lower bound is exactly same as the previously obtained lower bound as $p_1 + p_2 = 1$ satisfies $p_1p_2 = q_1q_2.$

It may be an interesting question that under which conditions $\kappa$ can attain -1 and +1. The upper correlation bound can be 1 iff $p_1(1-p_2) = (1-p_1)p_2, \quad p_2(1-p_1) = p_1(1-p_2) \Longleftrightarrow p_1 = p_2.$ So, $\kappa = \rho = 1$ iff $p_1 = p_2.$\\
The lower correlation bound can be -1 iff $p_1p_2 = (1-p_1)(1-p_2) \Longleftrightarrow p_1 + p_2 = 1.$ Now, by using $|\rho| \geq |\kappa|,$ we can claim that $\kappa = -1$ iff $p_1 = p_2 = \frac{1}{2}.$ So, only if all marginal probabilities are $\frac{1}{2}$, $\kappa$ can vary between -1 and 1, otherwise true bounds are narrower than that for binary case.

\section{Methods to generate Bivariate data with specified $\kappa$ and marginal probabilities}
\subsection{Using sorting based approach by Demirtas 2019}
Demirtas (2019) developed a sorting based algorithm to incorporate any level of feasible correlation in independently generated bivariate data. A similar algorithm can be followed to generate bivariate nominal or ordinal data given marginal probabilities and $\kappa$,
\begin{itemize}
    \item[] \textbf{step 1:} Find sorting based bounds of Cohen's kappa for given marginal probabilities (see section 4.1) and check given $\kappa$ is within the bounds. If $\kappa$ is outside of those bounds this method fails.
    \item[] \textbf{step 2:} Generate the variables, say, $X$ and $Y$ independently with given marginal probabilities. 
    \item[] \textbf{step 3:} Let's denote the specified $\kappa$ as $\kappa_{spec}$ and the lower and upper bounds as $\kappa_L$ and $\kappa_U$ respectively. Then do the following,\\
    if $\kappa_{spec} \geq 0$, sort first $100\frac{\kappa_{spec}}{\kappa_U}\%$ of both $X$ and $Y$ in ascending or descending order and keep rest of the data unchanged.\\
    if $\kappa_{spec} < 0$, sort first $100\frac{\kappa_{spec}}{\kappa_L}\%$ of $X$ and $Y$ in reverse order (i.e. if $X$ is sorted in ascending order then $Y$ must be sorted in descending order or vice versa) and keep rest of the data unchanged.
\end{itemize}

This method can be applied for bivariate data with any number of categories per rater. The problem with this method is that the sorting based bounds are actually not the exact bound of $\kappa$ given marginal probabilities. For some choices of marginals, these bounds are much narrower than the exact bounds, even in some cases, the sorting based upper bound can be negative and lower bound can be positive, which is ridiculous (see section 4.1 for examples). As a results, this sorting based method fails to generate data with some values $\kappa_{spec}$ even if that $\kappa_{spec}$ is contained in exact bounds of $\kappa.$\\ Interestingly, this problem won't occur for binary case. In binary case, there exists a linear relationship between $\kappa$ and $\rho$ and this sorting based approach works perfectly for correlation as sorting based correlation bounds are very good approximation of exact correlation bounds. So, sorting based binary $\kappa$ bounds are also very good approximation of exact binary $\kappa$ bounds, which ensures that sorting based approach works perfectly for generating bivariate binary data. 

\subsection{Proposed Algorithm to Generate Bivariate Data with Specified $\kappa$ and Marginals}
Let's assume $K$ categories for both raters A and B, with marginal probabilities, $\mathbf{p^A} = (p_1^A, p_2^A,..., p_K^A)^T$ and $\mathbf{p^B} = (p_1^B, p_2^B,..., p_K^B)^T$ respectively. Similar to the sorting based method, at first ratings of rater A and B are generated independently satisfying the marginal probabilities. Now, the goal is to sort these independent ratings in such a way to incorporate specified $\kappa$. In other words, we need to construct a $K \times K$ table which satisfies specified $\kappa$ and marginal probabilities should be same as the marginal probabilities of generated ratings. Let, $\mathbf{p_{(g)}^A} = (p_{1(g)}^A, p_{2(g)}^A,..., p_{K(g)}^A)^T$ and $\mathbf{p_{(g)}^B} = (p_{1(g)}^B, p_{2(g)}^B,..., p_{K(g)}^B)^T$ be the marginal probabilities of generated ratings of rater A and rater B respectively. 
In the definition of $\kappa$, $p_c$ is fixed once the marginals are specified ($p_c = \sum \limits_{i = 1}^K p_{i}^Ap_{i}^B$), so, it is easy to calculate the required value of $p_0$ to satisfy specified $\kappa$ (required $p_0$, $p_0^{req} = (1 - p_c)\kappa + p_c$). So, we need to find the cell probabilities of the $K \times K$ table which satisfies the known marginal probabilities and the sum of diagonal probabilities equal to $p_0^{req}$. For better understanding, let us start with $K = 3$. Consider the following table,
where $X$ and $Y$ are two nominal or ordinal variables denoting ratings of rater A and B respectively,
\begin{center}
\begin{tabular}{|c|c|c|c|c|}
\hline 
 & $Y = 1$ & $Y = 2$ & $Y = 3$ & Marginal of $X$ \\ 
\hline 
$X = 1$ & $a_{11}$ & $a_{12}$ & $a_{13}$ & $p_{1(g)}^A$ \\ 
\hline 
$X = 2$ & $a_{21}$ & $a_{22}$ & $a_{23}$ & $p_{2(g)}^A$ \\ 
\hline 
$X = 3$ & $a_{31}$ & $a_{32}$ & $a_{33}$ & $p_{3(g)}^A$ \\ 
\hline 
Marginal of $Y$ & $p_{1(g)}^B$ & $p_{2(g)}^B$ & $p_{3(g)}^B$ & 1 \\ 
\hline 
\end{tabular} 
\end{center}
We need to solve $\{a_{ij} : i = 1, 2, 3 \quad j = 1, 2, 3\}$ for non-negative roots, where, 
\begin{align*}
    a_{11} + a_{12} + a_{13} &= p_{1(g)}^A\\
    a_{21} + a_{22} + a_{23} &= p_{2(g)}^A\\
    a_{31} + a_{32} + a_{33} &= p_{3(g)}^A\\
    a_{11} + a_{21} + a_{31} &= p_{1(g)}^B\\
    a_{12} + a_{22} + a_{32} &= p_{2(g)}^B\\
    a_{13} + a_{23} + a_{33} &= p_{3(g)}^B\\
    a_{11} + a_{22} + a_{33} &= p_0^{req}.\\
\end{align*}
We can write this problem as to solve $\bm{A}^{7 \times 9}\mathbf{\bm{x}}^{9 \times 1} = \mathbf{\bm{b}}^{7 \times 1}$ under the constraint $\mathbf{\bm{x}} \geq \bm{0}, \quad \quad \quad \quad ... \quad \quad (\star)$\\
where,\\
\[ \bm{A} =
\begin{pmatrix}
1 & 1 & 1 & 0 & 0 & 0 & 0 & 0 & 0 \\
0 & 0 & 0 & 1 & 1 & 1 & 0 & 0 & 0 \\
0 & 0 & 0 & 0 & 0 & 0 & 1 & 1 & 1 \\
1 & 0 & 0 & 1 & 0 & 0 & 1 & 0 & 0 \\
0 & 1 & 0 & 0 & 1 & 0 & 0 & 1 & 0 \\
0 & 0 & 1 & 0 & 0 & 1 & 0 & 0 & 1 \\
1 & 0 & 0 & 0 & 1 & 0 & 0 & 0 & 1 \\
\end{pmatrix}
,\quad \quad \quad
\mathbf{\bm{x}} = 
\begin{pmatrix}
a_{11}\\
a_{12}\\
a_{13}\\
a_{21}\\
a_{22}\\
a_{23}\\
a_{31}\\
a_{32}\\
a_{33}\\
\end{pmatrix}
,\quad \quad \quad
\mathbf{\bm{b}} = 
\begin{pmatrix}
p_{1(g)}^A\\
p_{2(g)}^A\\
p_{3(g)}^A\\
p_{1(g)}^B\\
p_{2(g)}^B\\
p_{3(g)}^B\\
p_0^{req}\\
\end{pmatrix}
.
\]
This problem can be converted to the following optimization problem,\\
minimize $\bm{1}^T\mathbf{w}$ under the constraints, $\mathbf{x} \geq \bm{0}$, $\mathbf{w} \geq \bm{0}$ and $\bm{A}\mathbf{x} + \bm{I}\mathbf{w} = \mathbf{b}, \quad \quad \quad \quad \quad \quad \quad \quad \quad ... \quad \quad \quad (\star\star)$\\
where $\mathbf{w}$ is a vector of length 7, $\bm{1}$ is a unit vector of length 7, $\bm{I}$ is $7\times7$ identity matrix and $\bm{0}$ is null vector.
There exists a solution of $(\star)$ iff $\bm{1}^T\mathbf{w} = 0 \Longleftrightarrow \mathbf{w} = \bm{0}$ as $\mathbf{w} \geq \bm{0}$.\\
Now, let us summarize the algorithm for $K$ categories,
\begin{enumerate}[label= ]
    \item \textbf{Step 1:} Generate $n$ (specified) samples for rater A and rater B independently satisfying $\mathbf{p^A}$ \& $\mathbf{p^B}$ (both specified) and calculate generated marginal probabilities $\mathbf{p_{(g)}^A}$ \& $\mathbf{p_{(g)}^B}$. Calculate $p_0$ from specified $\mathbf{p^A}$ \& $\mathbf{p^B}$.
    \item \textbf{Step 2:} Formulate the problem as solving for $\mathbf{x}$ in $\bm{A}\mathbf{\bm{x}} = \mathbf{\bm{b}}$ under constraints $\mathbf{x} \geq \bm{0}$, where, $\bm{A}$ is $(2K + 1) \times K^2$ matrix constructed in the similar way as in the example, $\mathbf{x}$ is $K^2 \times 1$ vector of cell probabilities constructed by rows of the $K\times K$ table, $\mathbf{b}^{(2K+1) \times 1} = (\mathbf{p^A_{(g)}}, \mathbf{p^B_{(g)}}, p_0^{req})^T$.
    \item \textbf{Step 3:} Reformulate the problem as the following optimization problem,\\
    solve for $\textbf{u}$, for minimizing $\textbf{a}^T\textbf{u}$ under the constraints, $\textbf{u} \geq \bm{0}$, $\bm{B}\textbf{u} = \textbf{b}$,
    $\quad ... \quad (\star \star \star)$\\
    where,\\
    \[
    \mathbf{a} = 
    \begin{pmatrix}
    \bm{0}^{K^2 \times 1}\\
    \bm{1}^{(2K+1) \times 1}\\
    \end{pmatrix}
    , \quad \quad \quad
    \mathbf{u} = 
    \begin{pmatrix}
    \mathbf{x}^{K^2 \times 1}\\
    \mathbf{w}^{(2K+1) \times 1}\\
    \end{pmatrix}
    , \quad \quad \quad
    \bm{B} = 
    \begin{pmatrix}
    \bm{A}^{(2K+1) \times K^2} & \bm{I}^{(2K+1) \times (2K+1)} \\
    \end{pmatrix}
    .
    \]
    This is more compact version of $(\star\star)$.
    This optimization problem can be solved using simplex method. We have followed the algorithm described in chapter 8 of the book "Numerical Linear Algebra and Optimization Vol.1" by Gill, Murray, Wright, 1991.
    \item \textbf{Step 4:} Multiply the cell probabilities by $n$ to get the cell frequencies. If any cell frequency is not an integer at this point, make that an integer by rounding off. Due to rounding off, sum of the cell frequencies may not be $n$ by a very little margin; if it is less than $n$, increase a cell frequency, which was 0, by the appropriate amount or if it is greater than $n$, decrease the highest cell frequency by the appropriate amount. This adjustment is very small with respect to $n$, so it won't stop us to achieve specified $\kappa$ and marginal probabilities. 
    \item \textbf{Step 5:} The solved cell counts of the table denotes the frequencies of each pair of categories, so, these specify the generated bivariate data. For example, after step 4, a $2\times2$ table is found with cell counts of 4, 1, 3, 5 for cells (1, 1), (1, 2) , (2, 1), (2, 2) respectively, so in the generated data, there must be 4, 1, 3, 5 many pairs of categories (1, 1), (1, 2), (2, 1), (2, 2) respectively, so, generated data ((X, Y) pairs) will be (1, 1), (1, 1), (1, 1), (1, 1), (1, 2), (2, 1), (2, 1), (2, 1), (2, 2), (2, 2), (2, 2), (2, 2), (2, 2).

\end{enumerate}

\section{Methods to find exact bivariate bounds of $\kappa$ for given marginal probabilities}
\subsection{Sorting based bounds using the approach by Demirtas \& Hedeker 2011}
Demirtas \& Hedeker 2011 developed a sorting based method which approximately finds the true correlation bounds (bivariate) for given marginal distributions. Similar method can be obtained for finding $\kappa$ bounds, where, a large sample of random numbers are generated from both marginal distributions independently; the upper bound is approximated by the kappa between the bivariate sample after sorting the random numbers from both variables in same direction (ascending for both or descending for both) and the lower bound is approximated by the same but sorting is done in opposite direction (ascending for one and descending for the other or vice versa).\\

Due to linear relationship between correlation and kappa, sorting based bounds are very good approximation of exact kappa bounds for binary data. Unfortunately, when number of categories increases from 2, the sorting based bounds are much narrower than the exact kappa bounds for some marginal distributions. For example, consider the marginal probabilities (0.5, 0.4, 0.1) and (0.1, 0.4, 0.5) for two raters with categories 1, 2, 3; after sorting the ratings of both the raters in the same direction, we have first 10\% (1, 1) pairs, next 40\% (1, 2) pairs, next 40\% (2, 3) pairs and last 10\% (3, 3) pairs. Following table (2nd column) explains this for sample size 10,
\begin{center}
\begin{tabular}{|c|c|c|}
\hline 
Rater & Sorted sample (in same direction) & Sorted sample (in opposite direction) \\ 
\hline 
1 & 1, 1, 1, 1, 1, 2, 2, 2, 2, 3 & 1, 1, 1, 1, 1, 2, 2, 2, 2, 3 \\ 
\hline 
2 & 1, 2, 2, 2, 2, 3, 3, 3, 3, 3 & 3, 3, 3, 3, 3, 2, 2, 2, 2, 1\\ 
\hline 
\end{tabular} 
\end{center}
So, there is only 20\% agreement between the raters and resulting upper bound of $\kappa$ is $\frac{0.2 - (0.5\times0.1 + 0.4\times0.4 + 0.1\times0.5)}{1 - (0.5\times0.1 + 0.4\times0.4 + 0.1\times0.5)} \approx -0.08,$ whereas the exact upper bound (Cohen, 1960) is 0.46. If we sort the ratings in opposite direction (3rd column) we have 40\% agreement and resulting lower bound will be approximately 0.19, whereas the exact lower bound (section 4.2) is -0.35. So, using sorting based method, we have got a negative upper bound and the lower bound is greater than the upper bound, which is ridiculous. The sorting based bounds fails miserably when the marginal probability of the categories differs much, just like the last example.

\subsection{Proposed Algorithm to Find Exact Lower Bound of $\kappa$ given Marginals}
As we discussed earlier, in definition of $\kappa$, $p_c$ is fixed once marginal probabilities $\mathbf{p^A}$ and $\mathbf{p^B}$ (in the context of section 3.2) are specified. So, feasible values of $\kappa$ depends on feasible values of $p_0$. Given $\mathbf{p^A}$ and $\mathbf{p^B}$, Cohen (1960) found the upper bound by taking highest feasible value of $p_0$ as $p_0^{max} = \sum \limits_{i=1}^K min\{ p_i^A, p_i^B\}.$ One may think that $p_0$ is a probability measure, so, its lowest feasible value must be 0. This is true for some marginal probabilities but in general it is false. For example, consider $K=2$ with $\mathbf{p^A} = (.2, .8)^T$ and $\mathbf{p^B} = (.3, .7)^T$; let us try to construct $2 \times 2$ table satisfying these marginals and $p_0 = 0$ i.e. diagonal cell probabilities will be 0; observe that there does not exist a solution for off-diagonal entries to satisfy above marginals. So, finding the smallest feasible value of $p_0$ is a difficult problem indeed.\\

Our proposed algorithm to find the exact lower bound of $\kappa$ is based on the following idea. $p_0 = 0$ may not be feasible for some marginals but lowest feasible value of $p_0$ must be greater than or equal to 0 for any marginal. So, we can take 0 as the starting value of $p_0$ and start increasing $p_0$ from 0 until we get a feasible value of $p_0$ for given marginals. From section 3.2, we can claim that for a feasible value of $p_0$ there exist a solution for $(\star)$ i.e. $\mathbf{w} = \bm{0}$ will be the minimizer of $(\star\star)$. Note that, we must use $\mathbf{p^A}, \mathbf{p^B}$ in $\mathbf{b}$ instead of $\mathbf{p^A_{(g)}}, \mathbf{p^B_{(g)}}$ and this shows that the proposed algorithm is not simulation based, rather it is purely mathematical. Increasing $p_0$ from 0 may arise some questions like suitable step-size of the increment and there may be run-time issues if we take very small step sizes to achieve very good accuracy. So, we have used a bisection method to speed up the algorithm with better accuracy. At first we need to find two bounds of $p_0$ where, the lower bound is not a feasible value of $p_0$ but the upper bound is feasible. For some marginals if $p_0 = 0$ is not feasible, the easy choice of lower and upper bound is 0 and $p_0^{max}$ respectively. At each step we find the middle point of these bounds and check if that middle point is a feasible value of $p_0$; if feasible, then the middle point becomes the updated upper bound or if not feasible, then the middle point becomes he updated lower bound. This process will be continued until the desired accuracy is achieved and at the stopping point, we consider the updated upper bound as the exact lower bound of $p_0$. Let us summarize the algorithm in the following way,

\begin{itemize}[label =]
    \item \textbf{Step 1:} Set $p_0 = 0.$ Try to solve $(\star\star)$. If $\mathbf{w} = \bm{0}$ is the minimizer, then $\frac{ - \sum \limits_{i=1}^K p_i^A p_i^B}{1 - \sum \limits_{i=1}^K p_i^A p_i^B}$ is the exact lower bound of $\kappa$ given $\mathbf{p^A}$ and $\mathbf{p^B}$, otherwise go to step 2.
    \item \textbf{Step 2:} Set $lb = 0$ and $ub = \sum \limits_{i=1}^K min\{ p_i^A, p_i^B\}.$ Find $mb = \frac{lb + ub}{2}.$
    \item \textbf{Step 3:} Solve $(\star\star)$ by putting $p_0 = mb$ (see step 3 of the algorithm in section 3.2). If $\mathbf{w} = \bm{0}$ is the minimizer, set $ub = mb$, otherwise, set $lb = mb$ and again find $mb = \frac{lb + ub}{2}.$
    \item \textbf{Step 4:} Repeat step 3 until $\frac{ub-lb}{lb} < \epsilon$, where $\epsilon$ is very small specified quantity. At stopping point, denote $ub$ as $p_0^{min}$ and exact lower bound of $\kappa$ is found as $\frac{p_0^{min} - \sum \limits_{i=1}^K p_i^A p_i^B}{1 - \sum \limits_{i=1}^K p_i^A p_i^B}.$
\end{itemize}

 In section 2.2, we found the mathematical expression for bounds of $\kappa$
in binary case. Table 1 shows how accurately our proposed algorithm can find the exact lower bound as it matches exactly with the mathematical lower bound. As discussed in section 4.1, sorting based bounds can approximate the exact $\kappa$ bounds very well. We can see that in table 1.
\begin{table}[]
\caption{Binary Case: comparison among the formula based lower \& upper bound of $\kappa$ found in section 2.2, the sorting based bounds and our proposed lower bound ($\epsilon = 10^{-5}$) \& Cohen's upper bound with marginal probabilities $\mathbf{p^A} = (p_1, 1-p_1)^T$ and $\mathbf{p^B} = (p_2, 1-p_2)^T.$ All the values has been rounded up to five decimal place.}
\fontsize{9pt}{14pt}\selectfont
    \centering
    \begin{center}
    \begin{tabular}{|c|c|c|c|}
\hline \hline 
 ($p_1$, $p_2$) & \begin{tabular}{@{}c@{}}Formula Based Bounds \\ (section 2.2)\end{tabular} & \begin{tabular}{@{}c@{}}Sorting Based Bounds \\ (section 4.1)\end{tabular} & \begin{tabular}{@{}c@{}}Proposed lower bound \& Cohen's upper bound \\ (section 4.2)\end{tabular} \\ 
 \hline
 (0.1, 0.9) & (-0.21951, 0.02439) & (-0.22009, 0.02453) & (-0.21951, 0.02439) \\ 
 (0.2, 0.8) & (-0.47059, 0.11765) & (-0.46972, 0.11729 ) & (-0.47059, 0.11765) \\ 
 (0.3, 0.7) & (-0.72414, 0.31034) & (-0.72544, 0.31181 ) & (-0.72414, 0.31034) \\ 
 (0.4, 0.8) & (-0.92308 0.61538) & (-0.92125, 0.61566) & (-0.92308 0.61538) \\ 
 (0.5, 0.5) & (-1, 1) & (-0.99985, 0.99879) & (-1, 1) \\ 
\hline \hline
\end{tabular}
\end{center}

    \label{tab:my_label}
\end{table}

\begin{table}[]
\caption{Multi-category Case: comparison among following three types of bounds: (i) Sorting Based Bd : Sorting based bounds discussed in section 4.1,
(ii) Exact Bd: Proposed lower bound discussed in section 4.2 and Cohen's upper bound, (iii) $p_0 = 0$ Motivated Bd: Lower bound by putting $p_0 = 0$ in $\kappa$ definition, i.e. $\frac{ - \sum \limits_{i=1}^K p_i^A p_i^B}{1 - \sum \limits_{i=1}^K p_i^A p_i^B}$ and Cohen's upper bound i.e. $\frac{\sum \limits_{i=1}^K min\{ p_i^A, p_i^B\} - \sum \limits_{i=1}^K p_i^A p_i^B}{1 - \sum \limits_{i=1}^K p_i^A p_i^B}.$}
\fontsize{8.5pt}{14pt}\selectfont
    \centering
    \begin{tabular}{|c|c|c|c|c|c|}
    \hline \hline
      $K$ & $(\mathbf{p^A})^T$ & $(\mathbf{p^B})^T$ & Sorting Based Bd & Exact Bd & $p_0 = 0$ Motivated Bd\\
      \hline
      2 & (0.8, 0.2) & (0.7, 0.3) & (-0.3149, 0.7375) & (-0.3158, 0.7368) & (-1.6316, 0.7368)\\
      2 & (0.6, 0.4) & (0.4, 0.6) & (-0.9231, 0.6157) & (-0.9231, 0.6154) & (-0.9231 0.6154)\\
      3 & (0.8, 0.15, 0.05) & (0.8, 0.1, 0.1) & (-0.1765,  0.8517) & (-0.1765, 0.8529) & (-1.9412, 0.8529)\\
      3 & (0.8, 0.15, 0.05) & (0.05, 0.15, 0.8) & (0.0528, -0.0028) & (-0.1142, 0.1643) & (-0.1142, 0.1643)\\
      4 & (0.7, 0.1, 0.15, 0.05) & (0.7, 0.2, 0.05, 0.05) & (-0.2501, 0.7909) & (-0.2500, 0.7917) & (-1.0833, 0.7917)\\
      4 & (0.7, 0.1, 0.15, 0.05) & (0.05, 0.15, 0.1, 0.7) & (-0.1098,  0.0000) & (-0.1111, 0.2222) & (-0.1111, 0.2222)\\
      5 & (0.6, 0.1, 0.1, 0.1, 0.1) & (0.5, 0.1, 0.2, 0.1, 0.1) & (-0.3332, 0.9975) & (-0.3333, 1) & (-0.6667, 1)\\
      5 & (0.6, 0.1, 0.1, 0.1, 0.1) & (0.1, 0.1, 0.1, 0.1, 0.6) & (-0.0597, 0.0591) & (-0.1765, 0.4118) & (-0.1765, 0.4118)\\
      \hline \hline
    \end{tabular}
\end{table}

\subsection{Comparison Among Proposed Lower Bound \& Cohen's Upper Bound, Sorting Based Bounds and Bounds Motivated by $p_0 = 0$}
In section 4.1 and 2.2, we have discussed that for binary case sorting based bounds are good approximate of true bounds (see table 1 and first, second row of table 2) and $p_0 = 0$ is feasible only when $p_1 + p_2 = 1,$ so lower bound can be found by simply putting 0 for $p_0$ (2nd row of table 2). It is interesting to see under what conditions sorting based bounds and lower bound by $p_0 = 0$ works as good as true bounds. In this discussion, true lower bound is found by the algorithm we proposed in section 4.2 and the true upper bound is the upper bound found by Cohen (1960).\\

When $\mathbf{p^A}$ and $\mathbf{p^B}$ are component-wise close, sorting based bounds are very good approximation of true bounds but the lower bound found by $p_0 = 0$ is far away from the true lower bound. In table 2, examples of the above is shown for $K = 3, 4, 5$ (odd rows). The lower bound by $p_0 = 0$ is ridiculous (lesser than -1) under the above scenario when there is small number of categories but when the number of categories ($K$) increases the lower bound increases to above -1 but far away from the true lower bound. In the opposite scenario, when 
$\mathbf{p^A}$ and $\mathbf{p^B}$ are component-wise far, sorting based upper and lower bounds are much narrower than the true bounds but the lower bound by $p_0 = 0$ is accurate i.e. $p_0 = 0$ is feasible for this choice of marginals. In table 2, for $K = 3, 4, 5$ (even rows), the examples of this scenario is shown where the sorting based bounds are ridiculously bad as for $K = 3$ lower bound is positive while upper bound is negative, for $K=5$ sorting based bounds are too narrow but $p_0 = 0$ is feasible.\\ 

The above two scenarios are extreme and do not occur often in practice. In table 3, the marginal probabilities are generated randomly and independently (For $K$ categories, $K$ samples are generated from uniform (0, 1) and each of them is divided by their sum to get marginal probabilities) and for each choice of $K$, the percent of time $p_0 = 0$ produces the exact lower bound is presented in the table for $10^5$ replications. Interestingly, this percentage is only 70\% for $K=3$ but it is increasing to almost 100\% as $K$ increases. The following argument may serve as an intuition of this phenomena.\\
$p_0 = 0$ is feasible iff $\mathbf{w} = \bm{0}$ is a solution of $(\star \star).$ In step 3 of the algorithm in section 3.2, $(\star \star)$ is reformulated as $(\star \star \star)$, so, the above statement can be rephrased as, $p_0 = 0$ is feasible iff last $(2K + 1)$ components of $\mathbf{u}$ are 0. In chapter 8 of the book "Numerical Linear Algebra and Optimization Vol.1" by Gill, Murray, Wright, 1991, it is proved that in the simplex problem $(\star \star \star)$ (this kind of problems are called standard form simplex problem), if the rows of $\bm{B}$ are linearly independent, then the columns of $\bm{B}$ corresponding to positive elements in the minimizer $\mathbf{u}$ must be linearly independent. It can be proved that the rows of $\bm{B}$
are linearly independent. If $p_0 = 0$ has to be feasible, then the diagonal elements of the table have to be 0, so minimum number of zero components required in $\mathbf{u}$ is $(2K + 1) + K = 3K + 1.$ As the rows of $\bm{B}$ are linearly independent, $(2K+1)$ many columns of $\bm{B}$ must be linearly independent, so size of the largest subset of columns which are linearly dependent can be $K^2$ which implies minimum number of zero components in the minimizer is $K^2.$ Now, $K^2 > 3K + 1 \Longleftrightarrow K > 3$ as $K$ is positive integer. So, when number of categories are more than 3 per rater, there is a huge possibility that $p_0 = 0$ is feasible and this possibility will increase as number of categories increase.

\begin{table}[]
\caption{Percentage of cases (in $10^5$ replications) where $p_0 = 0$ gives the exact lower bounds while number of categories varies from 2 to 8 and marginals are generated randomly (satisfying sum of the probabilities is 1)}
\fontsize{10pt}{14pt}\selectfont
    \centering
    \begin{tabular}{|c|c|c|c|c|c|c|c|}
    \hline \hline
    $K$ & 3 & 4 & 5 & 6 & 7 & 8\\
    \hline
    Percentage & 69.013 & 94.668 & 98.737 & 99.416 & 99.692 & 99.839\\
    \hline \hline
    \end{tabular}
    \label{tab:my_label}
\end{table}

\section{Generating Multivariate Data with given kappa matrix and marginal probabilities}
There are several tools to generate multivariate binary data given correlation matrix and marginals for example, Demirtas 2006, Amatya \& Demirtas 2015 and so on. Due to the linear relationship between correlation and kappa (section 2.1) it is easy to find a unique correlation matrix from specified kappa matrix and data generation can be done using existing method. When number of categories is more than for each rater, we have described in section 3.1 that sorting based method by Demirtas (2019) will work but for narrower bounds (sorting based bounds). Multivariate extension of sorting based method is a future work and still in progress. In this section, we have extended our proposed algorithm to generate bivariate nominal or ordinal data (discussed in section 3.2) to the multivariate case, where the kappa matrix and marginal probabilities must be specified. We have assumed that the specified kappa matrix is feasible for given marginals if it is positive definite and each off-diagonal entry is within bivariate kappa bounds where lower bound is found by our proposed algorithm (section 4.2) and upper bound is found by Cohen (1960). We don't have any proof for the requirement of positive definiteness of kappa matrix but our motivation comes from the requirement of association matrices to be positive definite. The algorithm of generating multivariate data is exactly same as the algorithm for the bivariate case but the only difference is that instead of $K \times K$ table we need to use a multidimensional array. For bivariate case, finding matrix $\bm{A}$ from the $K \times K$ table was pretty straight forward but for multivariate case finding $\bm{A}$ from a multidimensional array may be little tricky. So, we have used the following algorithm in case of a $d$ dimensional array ($K$ categories per dimension) to construct the matrix $\bm{A}$,

\begin{enumerate}[label=]
    \item \textbf{Step 1:} Make a matrix $MAT$ of dimension $K^d \times d$, where each row of $MAT$ denotes a unique position in the $d$ dimensional array. The $i$th column of the matrix can be constructed as, 
    $r_{2i}$ many replications of the column vector\\ $(r_{1i}$ many 1, $r_{1i}$ many 2,..., $r_{1i}$ many $K)^T,$ where, $r_{1i} = K^{i-1}$ and $r_{2i} = K^{d-i}$, $i = 1, 2,..., d.$\\
    For example, when $K = 2$ and $d = 3$, $MAT = 
    \begin{pmatrix}
    1 & 1 & 1\\ 
    2 & 1 & 1\\
    1 & 2 & 1\\
    2 & 2 & 1\\
    1 & 1 & 2\\
    2 & 1 & 2\\
    1 & 2 & 2\\
    2 & 2 & 2\\
    \end{pmatrix}.$
    
    \item \textbf{Step 2:} Initialize matrix $\bm{A}$ as zero matrix. Dimension of $\bm{A}$ must be $[Kd + {d \choose 2}] \times K^d.$ Each of first $Kd$ rows is dedicated to the marginal probability of a category for a variable. For $j$th category of $i$th variable, find out the rows of $i$th column of $MAT$ which are equal to $j$. Lets denote the set of these rows as $i_j$ and the sub-matrix of $MAT$ with these rows as $MAT[i_j, ]$. These rows denote those cells in the $d$ dimensional array ($K$ categories per dimension) which corresponds to $j$th category of $i$th variable. Create a $d$ dimensional array with $K$ categories per dimension where the counts of the cells, whose positions are denoted by rows of $MAT[i_j, ]$, are 1 and other cell counts are 0. Collapse this array in the standard way to a vector and replace the $[(i-1)K + j]$th row of $\bm{A}$ with this vector. Do this for $j = 1, 2,..., K$ nested in each $i = 1, 2, ..., d$ to complete this step.
    
    \item \textbf{Step 3:} We are done with the rows corresponds to marginal probabilities. Now we need to create rows corresponds to sum of the diagonal cells for each pair of variables. For the pair $(i, j)$, find the rows of $MAT$ where elements of columns $i$ and $j$ are equal. Lets denote the set of these rows as $i_j$ and the sub-matrix of $MAT$ with these rows as $MAT[i_j, ]$. These rows denote the cells in the $d$ dimensional array ($K$ categories per dimension) containing same categories for variable $i$ and $j.$
    Create a $d$ dimensional array with $K$ categories per dimension where the counts of the cells, whose positions are denoted by rows of $MAT[i_j, ]$, are 1 and other cell counts are 0. Collapse this array in the standard way to a vector and replace the $[Kd + s]$th row of $\bm{A}$ with this vector. Here $s$ denotes the position of $(i, j)$ in the set $S = \{(p, q): p = 1, 2,..., d; q = 1, 2,..., d; p<q\}$ where order of the elements of the set can be any but must be noted for future use.
\end{enumerate}

Once the matrix $\bm{A}$ is done, we need to generate $d$ variable independently with specified marginal probabilities to find the generated marginal probabilities. We can calculate the sum of the diagonal probabilities required for each pair of variables to attain the specified kappa value in the kappa matrix using the same formula described in section 3.2. Now, the vector $\mathbf{b}$ can be created maintaining the following ordering:\\
generated marginal probabilities of variable 1 with the marginals in order of categories $1, 2,..., K$, generated marginal probabilities of variable 2 with the marginals in order of categories $1, 2,..., K$,..., generated marginal probabilities of variable $K$ with the marginals in order of categories $1, 2,..., K$, sum of the diagonal probabilities for pair of variables maintaining same order in set S, which is described in step 3.\\
If matrix $\bm{A}$ and vector $\mathbf{b}$ are constructed, step 3, 4, 5 of bivariate data generation in section 3.2 can be followed to generate multivariate data with given kappa matrix and marginals. Note that in $(\star\star\star)$ $\mathbf{a}, \mathbf{u}, \bm{B}$ will be (only the dimensions are changed),\\ 
\[
    \mathbf{a} = 
    \begin{pmatrix}
    \bm{0}^{K^d \times 1}\\
    \bm{1}^{(Kd+{d \choose 2}) \times 1}\\
    \end{pmatrix}
    , \quad \quad \quad
    \mathbf{u} = 
    \begin{pmatrix}
    \mathbf{x}^{K^d \times 1}\\
    \mathbf{w}^{(Kd+{d \choose 2}) \times 1}\\
    \end{pmatrix}
    , \quad \quad \quad
    \bm{B} = 
    \begin{pmatrix}
    \bm{A}^{(Kd+{d \choose 2}) \times K^d} & \bm{I}^{(Kd+{d \choose 2}) \times (Kd+{d \choose 2})} \\
    \end{pmatrix}
    .
    \]
    
Table 4 shows that this method allows us to generate multivariate data where generated marginals and kappa matrix match with the specified ones very accurately, even with a moderate sample size. \\

Unfortunately, this method has limitations. This limitation is not theoretical but computational. For multivariate case, we are dealing with a $d$ dimensional array where number of elements are $K^d$ and the number of elements in $\mathbf{a}, \mathbf{u}, \bm{B}$ are $K^d + Kd + {d \choose 2}$, $K^d + Kd + {d \choose 2}$ and $(Kd + {d \choose 2})(K^d + Kd + {d \choose 2})$ respectively. These quantities increase rapidly with the increase of $d$. Although theoretically, this algorithm works for any choices of $d$ and $K$, in practice, this algorithm becomes computationally very heavy when $K^d$ is large. So, while using the function \textit{rmvNomOrd} in the R-package \textit{mvtNomOrd} (see section 6), we advise users to keep the number of variables less than or equal to 18, 11, 9, 8, 8 where number of categories per variable is 2, 3, 4, 5, 6 respectively.

\begin{sidewaystable}[]
    \caption{Comparison of generated PMAT and KMAT with specified PMAT and  KMAT for $d = 2, 3, 4, 5$ where 1000 rows of data generated in each case; PMAT : marginal probability matrix where ith column denotes the marginal of ith variable; KMAT : Cohen's kappa matrix of dimension $d$. For this table $K=4$ for all variables.}
    \fontsize{8pt}{14pt}\selectfont
    \centering
    \begin{tabular}{|c|c|c|c|}
    \hline \hline
      Specified PMAT & Generated PMAT & Specified KMAT & Generated KMAT \\
      $\begin{pmatrix}
     0.16 & 0.27\\
     0.29 & 0.33\\
     0.29 & 0.07\\
     0.26 & 0.33\\
      \end{pmatrix}$
      &
      $\begin{pmatrix}
     0.148 & 0.282\\
     0.283 & 0.314\\
     0.301 & 0.078\\
     0.268 & 0.326\\
      \end{pmatrix}$
      &
      $\begin{pmatrix}
      1 & 0.65\\
      0.65 & 1\\
      \end{pmatrix}$
      &
    $\begin{pmatrix}
      1 & 0.652\\
      0.652 & 1\\
    \end{pmatrix}$\\
    \hline
      $\begin{pmatrix}
    0.21 & 0.17 & 0.51\\
    0.26 & 0.23 & 0.14\\
    0.16 & 0.30 & 0.21\\
    0.37 & 0.30 & 0.14\\
      \end{pmatrix}$
      &
      $\begin{pmatrix}
    0.208 & 0.177 & 0.523\\
    0.267 & 0.228 & 0.133\\
    0.152 & 0.303 & 0.210\\
    0.373 & 0.292 & 0.134\\
      \end{pmatrix}$
      &
      $\begin{pmatrix}
       1 & -0.11 & 0.32\\
     -0.11 & 1 & 0.11\\
      0.32 & 0.11 & 1\\
      \end{pmatrix}$
      &
      $\begin{pmatrix}
     1 & -0.108 & 0.323\\
     -0.108 & 1 & 0.110\\
     0.323 & 0.11 & 1\\
      \end{pmatrix}$
    \\
    \hline
    $\begin{pmatrix}
    0.32 & 0.19 & 0.099 & 0.13\\
    0.12 & 0.38 & 0.475 & 0.43\\
    0.28 & 0.31 & 0.188 & 0.35\\
    0.28 & 0.12 & 0.238 & 0.09\\
    \end{pmatrix}$
    &
    $\begin{pmatrix}
    0.304 & 0.194 & 0.100 & 0.115\\
    0.102 & 0.399 & 0.472 & 0.447\\
    0.299 & 0.292 & 0.195 & 0.326\\
    0.295 & 0.115 & 0.233 & 0.112\\
    \end{pmatrix}$
    &
    \begingroup
    \setlength\arraycolsep{1.5pt}
    $\begin{pmatrix}
    1 & 0.39 & -0.24 & 0.32\\
    0.39 & 1 & 0.05 & -0.04\\
    -0.24 & 0.05 & 1 & 0.31\\
    0.32 & -0.04 & 0.31 & 1\\
    \end{pmatrix}$
    \endgroup
    &
    \begingroup
    \setlength\arraycolsep{1.5pt}
    $\begin{pmatrix}
   1 & 0.393 & -0.236 & 0.324\\
   0.393 & 1 & 0.046 & -0.042\\
  -0.236 & 0.046 & 1 & 0.303\\
   0.324 & -0.042 & 0.303 & 1\\
    \end{pmatrix}$
    \endgroup
    \\
    \hline
    $\begin{pmatrix}
    0.25 & 0.35 & 0.19 & 0.287 & 0.293\\
    0.17 & 0.20 & 0.24 & 0.168 & 0.293\\
    0.25 & 0.30 & 0.43 & 0.376 & 0.263\\
    0.33 & 0.15 & 0.14 & 0.169 & 0.151\\
    \end{pmatrix}$
    &
    $\begin{pmatrix}
    0.248 & 0.356 & 0.180 & 0.273 & 0.278\\
    0.154 & 0.181 & 0.251 & 0.177 & 0.294\\
    0.262 & 0.312 & 0.439 & 0.373 & 0.271\\
    0.336 & 0.151 & 0.130 & 0.177 & 0.157\\
    \end{pmatrix}$
    &
    \begingroup
    \setlength\arraycolsep{1.3pt}
    $\begin{pmatrix}
     1 & 0.14 & -0.22 & -0.07 & 0.39\\
     0.14 & 1 & -0.12 & -0.12 & -0.26\\
    -0.22 & -0.12 & 1 & 0.21 & -0.18\\
    -0.07 & -0.12 & 0.21 & 1 & -0.11\\
     0.39 & -0.26 & -0.18 & -0.11 &  1\\
    \end{pmatrix}$
    \endgroup
    &
    \begingroup
    \setlength\arraycolsep{1.3pt}
    $\begin{pmatrix}
     1 & 0.139 & -0.222 & -0.075 & 0.39\\
    0.139 & 1 & -0.120 & -0.119 & -0.254\\
    -0.222 & -0.120 & 1 & 0.213 & -0.184\\
    -0.075 & -0.119 & 0.213 & 1 & -0.108\\
    0.39 & -0.254 & -0.184 & -0.108 & 1\\
    \end{pmatrix}$
    \endgroup
    \\
    \hline \hline     
    \end{tabular}
\end{sidewaystable}

\section{The R-Package: \lowercase{mvt}N\lowercase{om}O\lowercase{rd}}
We have written an R-package \textit{mvtNomOrd} to implement the algorithms we have proposed in this paper.
This package consists of four functions. The function \textit{cohen\_kappa} calculates Cohen’s kappa measure for given $K \times K$ table or bivariate data using the definition of $\kappa$.
The function \textit{find\_kappa\_bound} finds the exact lower bound of Cohen’s kappa based on the method described in section 4.2, and the exact upper bound is found based on Cohen (1960), for given marginal probabilities. The function \textit{validate} validates and adjusts the user-inputted marginal probabilities of each nominal or ordinal variables for chosen number of categories. If they are fine, the function proceeds to validate the kappa matrix for chosen number of dimensions. Our main function \textit{rmvNomOrd} generates multivariate nominal and ordinal data with user-specified kappa matrix and marginal probabilities using the method described in section 5.

\section{References}
\begin{enumerate}
    \item  Cohen, Jacob (1960). ``A coefficient of agreement for nominal scales", \textit{Educational and Psychological Measurement}, 20 (1), 37–46. 
    \item  McHugh, Mary L. (2012).``Interrater reliability: The kappa statistic", \textit{Biochemia Medica}, 22 (3), 276–282.
    \item Fleiss, J.L. (1971). ``Measuring nominal scale agreement among many raters", \textit{Psychological Bulletin}. 76 (5), 378–382.
    \item Demirtas, H. (2006), ``A Method for Multivariate Ordinal Data Generation Given Marginal Distributions and Correlations”, \textit{Journal of Statistical Computation and Simulation}, 76, 1017–1025.
    \item Anup Amatya \& Hakan Demirtas (2015) MultiOrd: An R Package for Generating Correlated Ordinal Data, \textit{Communications in Statistics - Simulation and Computation}, 44:7, 1683-1691.
    \item Hakan Demirtas (2019) Inducing Any Feasible Level of Correlation to Bivariate Data With Any Marginals, \textit{The American Statistician}, 73:3, 273-277.
    \item Demirtas, H., and Hedeker, D. (2011),``A Practical Way for Computing Approximate Lower and Upper Correlation Bounds”, \textit{The American Statistician}, 65, 104–109.
    \item Emrich, L. J., and Piedmonte, M. R. (1991),``A Method for Generating HighDimensional Multivariate Binary Variates”, \textit{The American Statistician}, 45, 302–304.
    \item Gill, P.E., Murray, W. and Wright, M.H., ``The Simplex Method" in \textit{Numerical Linear Algebra and Optimization Vol. 1}, Addison-Wesley, 1991, ch. 8, pp. 337-407.
\end{enumerate}

\end{document}